# SIMULTANEOUS INFERENCE: WHEN SHOULD HYPOTHESIS TESTING PROBLEMS BE COMBINED?

By Bradley Efron

*Stanford University*

Modern statisticians are often presented with hundreds or thousands of hypothesis testing problems to evaluate at the same time, generated from new scientific technologies such as microarrays, medical and satellite imaging devices, or flow cytometry counters. The relevant statistical literature tends to begin with the tacit assumption that a single combined analysis, for instance, a False Discovery Rate assessment, should be applied to the entire set of problems at hand. This can be a dangerous assumption, as the examples in the paper show, leading to overly conservative or overly liberal conclusions within any particular subclass of the cases. A simple Bayesian theory yields a succinct description of the effects of separation or combination on false discovery rate analyses. The theory allows efficient testing within small subclasses, and has applications to "enrichment," the detection of multi-case effects.

**1. Introduction.** Modern scientific devices such as microarrays routinely provide the statistician with thousands of hypothesis testing problems to consider at the same time. A variety of statistical techniques, false discovery rates, family-wise error rates, permutation methods etc., have been proposed to handle large-scale testing situations, usually under the tacit assumption that all available tests should be analyzed together—for instance, employing a single false discovery analysis for all the genes in a given microarray experiment.

This can be a dangerous assumption. As my examples will show, omnibus combination may distort individual inferences in both directions: highly significant cases may be hidden while insignificant ones are enhanced. This paper concerns the choice between combination and separation of hypothesis testing problems. A helpful methodology will be described for diagnosing when separation may be necessary for a subset of the testing problems, as well as for carrying out separation in an efficient fashion.

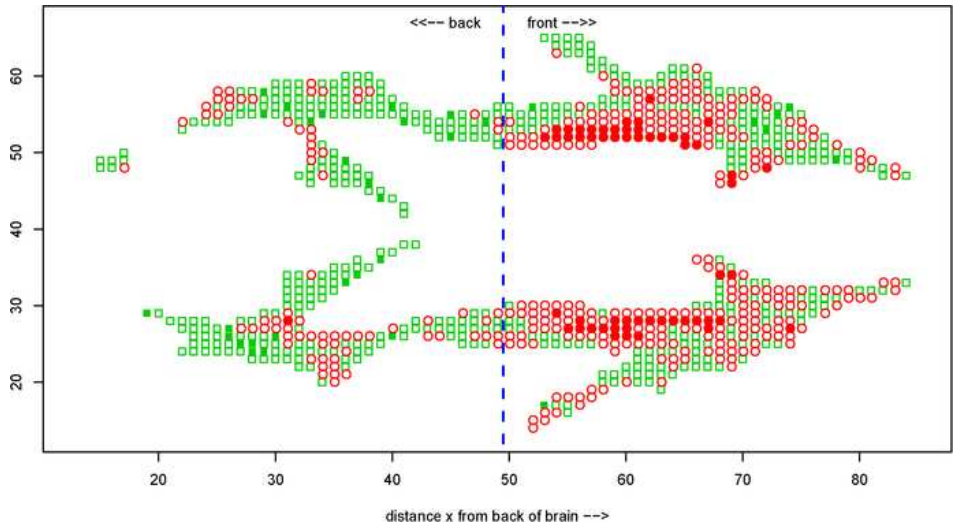

FIG. 1. *Brain Data: z-values comparing 6 dyslexic children with 6 normals; horizontal section showing 848 of 15443 voxels. Code: Red z ≥ 0, Green z < 0; solid circles z ≥ 2.0, solid squares z ≤ −2.0; "x" indicates distance from back of brain; y-axis is right-left distance. The front half of the brain appears to have more positive z-values. Data from Schwartzman, Dougherty and Taylor* (2005).

Figure 1 illustrates our motivating example, taken from Schwartzman, Dougherty and Taylor (2005). Twelve children, six dyslexic and six normal, received Diffusion Tensor Imaging (DTI) brain scans, an advanced form of MRI technology that measures fluid diffusion in the brain, in this case at $N = 15443$ locations, each represented by its own voxel's response. A $z$-value "$z_i$" comparing the dyslexics with the normals has been calculated for each voxel $i$, such that $z_i$ should have a standard normal distribution under the null hypothesis of no dyslexic-normal difference at that brain location,

$$(1.1) \qquad \textit{theoretical null hypothesis}: z_i \sim N(0, 1).$$

The $z$-values for a horizontal section of the brain containing 848 of the 15443 voxels are indicated in Figure 1. Distance "$x$" from the back toward the front of the brain is indicated along the $x$ axis. In appearance at least, the $z_i$'s seem to be more positive toward the front.

The investigators were, of course, interested in spotting locations of genuine brain differences between the dyslexic and normal children. A standard False Discovery Rate (FDR) analysis described in Section 2, Benjamini and Hochberg (1995), returned 198 "significant" voxels at threshold level $q = 0.1$, those having $z_i \geq 3.02$. The histogram of all 15443 $z_i$'s appears in the left panel of Figure 2.

Separate $z$-value histograms for the back and front halves of the brain are displayed in the right panel of Figure 2, with the dividing line at $x = 49.5$ as shown in Figure 1. Two discrepancies strike the eye: the heavy right tail seen in the combined histogram on the right comes almost exclusively from front-half voxels; and the center of the back-half histogram is shifted leftward about 0.35 units compared to the front.

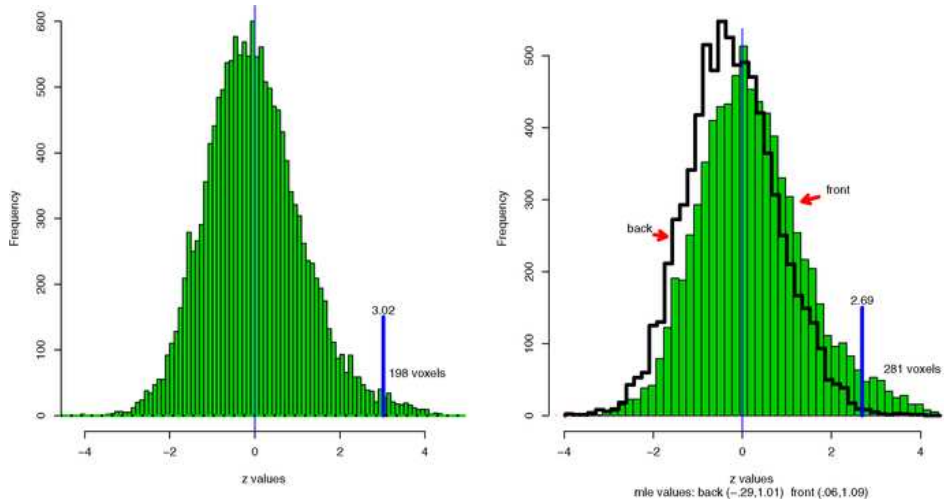

Fig. 2. *Left panel: histogram of all 15433 z-values for Brain Data; the 198 voxels with $z_i \geq 3.02$ were judged significant by an FDR analysis with threshold level $q = 0.1$. Right panel: histograms for back and front voxels; separate FDR analyses at level $q = 0.1$ gave no significant voxels for back half, and 281 significant voxels, those with $z_i \geq 2.69$, for front half. MLE values are means and standard deviations for normal densities fit to the centers of the two histograms, as explained in Section 3.*

Separate FDR analyses, each at threshold level $q = 0.1$, were run on the back and front half data. 281 front half voxels were found significant, those having $z_i \geq 2.69$; the back half analysis gave no significant voxels at all, in contrast to 9 significant back-half cases found in the combined analysis. This example illustrates both of the dangers of combination—over and under sensitivity within different subclasses of the experiment.

Section 2 begins with a simple Bayesian theorem that quantifies the choice between separate and combined analysis. It is applied to the brain data in Section 3, elucidating the differences between front and back false discovery rate analyses. The theorem is most useful for separately investigating *small* subclasses, where there is too little data for the direct empirical Bayes techniques of Section 3. Sections 4 and 7 demonstrate how *all* the data, all $N = 15833$ $z$ values in the Brain study, can be brought to bear on efficient FDR estimation for a small subclass, for example, just the 82 voxels at distance $x = 18$. Section 6 applies small subclass theory to "enrichment," the assessment of a possible overall discrepancy between the $z$-values within and outside a chosen class.

A reasonable objection to performing separate analyses on portions of the $N$ cases is the possibility of weakening control of Type 1 error, the overall size of the procedure. This question is taken up in Section 5, where false discovery rate methods are shown to be nearly immune to the danger. Some technical remarks and discussion end the paper in Sections 8 and 9.

The question of separating large-scale testing problems has not received much recent attention. Two relevant references are Genovese, Roeder and Wasserman (2006), and Ferkinstad, Frigessi, Thorleifsson and Kong (2007). Enrichment techniques have been more actively developed, as in Subramanian et al. (2005), Newton et al. (2007) and Efron and Tibshirani (2007).

**2. A separate-class model.** The "Two-Groups model," reviewed below, provides a simple Bayesian framework for the analysis of simultaneous hypothesis testing problems. The framework will be extended here to include the possibility of separate situations in different sub-classes of problems—for example, for the back and front halves of the brain in Figure 1—the extended framework being the "Separate-Class model."

First, we begin with a brief review of the Two-Groups model, taken from Efron (2005, 2007a). It starts with the Bayesian assumption that each of the $N$ cases (all $N = 15443$ voxels for the Brain Data) is either null or nonnull, with prior probability $p_0$ or $p_1 = 1 - p_0$, and with its test statistic "$z$" having density either $f_0(z)$ or $f_1(z)$,

$$(2.1) \quad \begin{array}{ll} p_0 = \mathrm{Prob}\{\mathrm{null}\}, & f_0(z) \text{ density if null,} \\ p_1 = \mathrm{Prob}\{\mathrm{nonnull}\}, & f_1(z) \text{ density if nonnull.} \end{array}$$

The theoretical null model (1.1) makes $f_0(z)$ a standard normal density

$$(2.2) \quad f_0(z) = \varphi(z) = \frac{1}{\sqrt{2\pi}} \; e^{-1/2 z^2},$$

an assumption we will question more critically later. We add the qualitative requirement that $p_0$ is large, say,

$$(2.3) \quad p_0 \geq 0.90,$$

reflecting the usual purpose of large-scale testing, which is to reduce a vast set of possibilities to a much smaller set of interesting prospects.

Model (2.1) is particularly helpful for motivating False Discovery Rate methods. Let $F_0(z)$ and $F_1(z)$ be the cumulative distribution functions (c.d.f.) corresponding to $f_0(z)$ and $f_1(z)$, and define the mixture c.d.f.

$$(2.4) \quad F(z) = p_0 F_0(z) + p_1 F_1(z).$$

Then the a posteriori probability of a case being in the null group of (2.1), given that its $z$-value $z_i$ is less than some threshold $z$, is the "Bayesian false discovery rate"

$$(2.5) \qquad \text{Fdr}(z) \equiv \text{Prob}\{\text{null} \,|\, z_i < z\} = p_0 F_0(z)/F(z).$$

[It is notationally convenient here to consider the negative end of the $z$-scale, e.g., $z = -3$, but we could just as well take $z_i > z$ or $|z_i| > z$ in (2.5).]

Benjamini and Hochberg's (1995) false discovery control rule estimates $\text{Fdr}(z)$ by

$$(2.6) \qquad \overline{\text{Fdr}}(z) = p_0 F_0(z)/\bar{F}(z),$$

where $\bar{F}(z)$ is the empirical c.d.f.

$$(2.7) \qquad \bar{F}(z) = \#\{z_i \leq z\}/N.$$

Having selected some control level "$q$," often $q = 0.1$, the rule declares all cases as nonnull having $z_i \leq z_{\max}$, where $z_{\max}$ is the maximum value of $z$ satisfying

$$(2.8) \qquad \overline{\text{Fdr}}(z_{\max}) \leq q.$$

[Usually taking $p_0 = 1$ and $F_0(z)$ the theoretical null c.d.f. $\Phi(z)$ in (2.5).]

Rule (2.8), which looks Bayesian here, can be shown to have an important frequentist "control" property: if the $z_i$'s are independent, the expected proportion of false discoveries, that is, the proportion of cases identified by (2.8) that are actually from the null group in (2.1), will be no greater than $q$. Benjamini and Yekutieli (2001) relax the independence requirement somewhat. Most large-scale testing situations exhibit substantial correlations among the $z$ values—obvious in Figure 1—but dependence is less of a problem for the empirical Bayes approach to false discovery rates we will follow here [see Efron (2007a, 2007b)].

Defining the *mixture density* $f(z)$ from (2.1),

$$(2.9) \qquad f(z) = p_0 f_0(z) + p_1 f_1(z)$$

leads to the "local false discovery rate" $\text{fdr}(z)$,

$$(2.10) \qquad \text{fdr}(z) \equiv \text{Prob}\{\text{null} \,|\, z_i = z\} = p_0 f_0(z)/f(z)$$

for the probability of a case being in the null group given $z$-score $z$. Densities are more natural than the tail areas of (2.5) for Bayesian analysis. Both will be used in what follows.

The *Separate-Class model* extends (2.1) to cover the situation where the $N$ cases can be divided into distinct classes, possibly having different choices of $p_0$, $f_0$ and $f_1$. Figure 3 illustrates the scheme: the two classes "A" and "B" (front and back in Figure 2) have a priori probabilities $\pi_A$ and $\pi_B = 1 - \pi_A$.

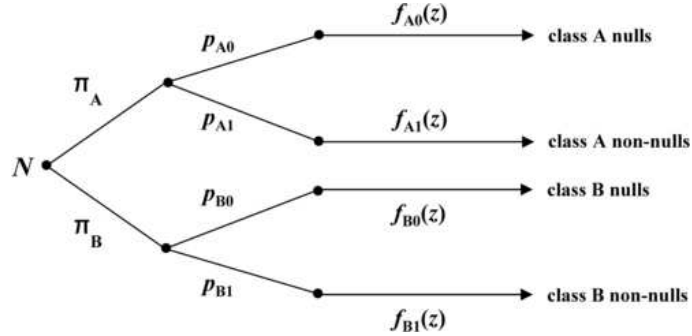

Fig. 3. *A Separate-Class model with two classes: The N cases separate into classes A or B with a priori probability $\pi_A$ or $\pi_B$; the Two-Groups model (2.1) holds separately, with possibly different parameters, within each class.*

The Two-Groups model (2.1) holds separately within each class, for example, with $p_0 = p_{A0}$, $f_0(z) = f_{A0}(z)$, and $f_1(z) = f_{A1}(z)$ in class A. It is important to notice that the class label, A or B, is observed by the statistician, in contrast to the null/nonnull dichotomy, which must be inferred.

Our previous definitions apply separately to classes A and B, for instance, following (2.9)–(2.10),

$$(2.11) \qquad \begin{aligned} f_A(z) &= p_{A0} f_{A0}(z) + p_{A1} f_{A1}(z) \quad \text{and} \\ \text{fdr}_A(z) &= p_{A0} f_{A0}(z)/f_A(z). \end{aligned}$$

Combining the two classes in Figure 3 gives marginal densities

$$(2.12) \qquad \begin{aligned} f_0(z) &= \pi_A f_{A0}(z) + \pi_B f_{B0}(z), \\ f_1(z) &= \pi_A f_{A1}(z) + \pi_B f_{B1}(z), \end{aligned}$$

and $p_0 = \pi_A p_{A_0} + \pi_B p_{B_0}$, so

$$(2.13) \qquad f(z) = \pi_A f_A(z) + \pi_B f_B(z),$$

leading to the combined local false discovery rate $\text{fdr}(z) = p_0 f_0(z)/f(z)$ as in (2.10). The same relationships with c.d.f.'s replacing densities apply to tail area Fdr's, (2.5).

Bayes theorem yields a simple but useful relationship between the separate and combined false discovery rates:

THEOREM. *Define $\pi_A(z)$ as the conditional probability of class A given $z$,*

$$(2.14) \qquad \pi_A(z) = \text{Prob}\{A|z\},$$

*and also*

$$(2.15) \qquad \pi_{A0}(z) = \mathrm{Prob}_0\{A|z\},$$

*the conditional probability of class $A$ given $z$ for a null case. Then*

$$(2.16) \qquad \mathrm{fdr}_A(z) = \mathrm{fdr}(z) \cdot \frac{\pi_{A0}(z)}{\pi_A(z)}.$$

PROOF. Let $I$ be the event that a case is null, so $I$ occurs in the two null paths of Figure 3 but not otherwise. Then, from definition (2.10),

$$
\begin{aligned}
(2.17) \qquad \frac{\mathrm{fdr}_A(z)}{\mathrm{fdr}(z)} &= \frac{\mathrm{Prob}\{I|A,z\}}{\mathrm{Prob}\{I|z\}} = \frac{\mathrm{Prob}\{I,A|z\}}{\mathrm{Prob}\{A|z\}\,\mathrm{Prob}\{I|z\}} \\
&= \frac{\mathrm{Prob}\{A|I,z\}}{\mathrm{Prob}\{A|z\}} = \frac{\pi_{A0}(z)}{\pi_A(z)}. \qquad \square
\end{aligned}
$$

The Theorem has useful practical applications. Section 4 shows that the ratio

$$(2.18) \qquad R_A(z) = \pi_{A0}(z)/\pi_A(z)$$

in (2.16) can often be easily estimated, yielding helpful diagnostics for possible discrepancies between $\mathrm{fdr}_A(z)$ and $\mathrm{fdr}(z)$, the separate and combined false discovery rates.

Tail-area false discovery rates (2.5) also follow (2.16), after the obvious definitional changes,

$$(2.19) \qquad \mathrm{Fdr}_A(z) = \mathrm{Fdr}(z) \cdot R_A(z),$$

where now $R_A(z)$ involves probabilities for cases having $z_i \leq z$,

$$(2.20) \qquad R_A(z) = \frac{\mathrm{Prob}_0\{A|z_i \leq z\}}{\mathrm{Prob}\{A|z_i \leq z\}}.$$

There is no real reason, except expositional clarity, for restricting attention to just two classes. Section 4 briefly discusses versions of the Theorem applicable to more nuanced situations—in terms of Figure 1, for example, where the relevance of other cases to the fdr at a given "$x$" falls off smoothly as we move away from $x$. First though, Section 3 applies the Theorem to the dichotomous front-back Brain Data analysis.

**3. Class-wise Fdr estimation.** The Theorem of Section 2 says that separate and combined local false discovery rates are related by

$$(3.1) \qquad \mathrm{fdr}_A(z) = \mathrm{fdr}(z) \cdot R_A(z), \qquad R_A(z) = \pi_{A0}(z)/\pi_A(z),$$

where $\pi_{A0}(z)$ and $\pi_A(z)$ are the conditional probabilities $\mathrm{Prob}_0\{A|z\}$ and $\mathrm{Prob}\{A|z\}$. This section applies (3.1) to the Brain Data of Figures 1 and 2, taking the front-half voxels as class A. The front-back dichotomy is extended to a more realistic multi-class model in Section 4.

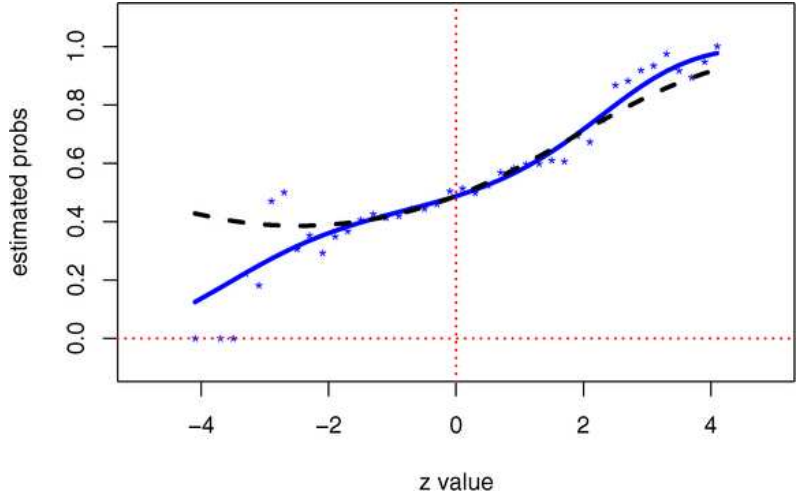

Fig. 4. *Points are proportion of front-half voxels $r_{Ak}$, (3.2), for Brain Data of Figure 2. Solid curve is $\hat{\pi}_A(z)$, cubic logistic regression estimate of $\pi_A(z) = \text{Prob}\{A|z\}$; Dashed curve $\hat{\pi}_{A0}(z)$ estimates $\text{Prob}_0\{A|z\}$ as explained in text.*

In order to estimate $\pi_A(z)$, it is convenient, though not necessary, to bin the $z$-values. Define $r_{Ak}$ as the proportion of class A $z$-values in bin $k$,

$$(3.2) \qquad r_{Ak} = N_{Ak}/N_k,$$

with $N_k$ the number of $z_i$'s observed in bin $k$, and $N_{Ak}$ the number of those originating from voxels in class A. The points in Figure 4 show $r_{Ak}$ for $K = 42$ bins, each of length 0.2, running from $z = -4.2$ to 4.2. As suggested by the right panel of Figure 2, the proportions $r_{Ak}$ steadily increase as we move from left to right.

A standard weighted logistic regression, fitting $\text{logit}(\pi_{Ak})$ as a cubic function of midpoint $z_{(k)}$ of bin $k$, with weights $N_k$, yielded estimate $\hat{\pi}_A(z)$ shown by the solid curve in Figure 4. Binning isn't necessary here, but it is comforting to see $\hat{\pi}_A(z)$ nicely following the $r_{Ak}$ points. (The three bins with $r_{Ak} = 0$ at the extreme left contain only $N_k = 2$ $z_i$'s each.)

In order to estimate $\pi_{A0}(z)$, we need to make some assumptions about the null distributions in the Two-Class model of Figure 3. Following Efron (2004a, 2004b), we assume that $f_{A0}(z)$ and $f_{B0}(z)$ are normal densities, but not necessarily $N(0, 1)$ as in (1.1), say,

$$(3.3) \qquad f_{A0}(z) \sim N(\delta_{A0}, \sigma_{A0}^2) \quad \text{and} \quad f_{B0}(z) \sim N(\delta_{B0}, \sigma_{B0}^2).$$

Bayes theorem then gives

$$\frac{\pi_{A0}(z)}{\pi_{B0}(z)} = \frac{\pi_{A0}(z)}{1 - \pi_{A0}(z)}$$

TABLE 1
*Parameter estimates for the null arms of the Two-Class model in Figure 3, brain data. Obtained using R program locfdr, Efron (2007b), MLE fitting method*

|            | $\widehat{p}_0$ | $\widehat{\delta}_0$ | $\widehat{\sigma}_0$ | $\pi$ |
|------------|------|-------|-------|------|
| A (front): | 0.97 | 0.06  | 1.09  | 0.50 |
| B (back):  | 1.00 | −0.29 | 1.01  | 0.50 |

$$(3.4) \qquad = \frac{\pi_A p_{A0} \sigma_{B0}}{\pi_B p_{B0} \sigma_{A0}} \exp\left\{-0.5\left[\left(\frac{z - \delta_{A0}}{\sigma_{A0}}\right)^2 - \left(\frac{z - \delta_{B0}}{\sigma_{B0}}\right)^2\right]\right\},$$

which is easily solved for $\pi_{A0}(z)$.

The R algorithm *locfdr* used "MLE fitting" described in Section 4 of Efron (2007b) to provide the parameter estimates in Table 1. (The front-back dividing line in Figure 1 was chosen to put about half the voxels into each class, so $\pi_A / \pi_B \doteq 1$.) Solving for $\widehat{\pi}_{A0}$ in (3.4) gave the dashed curve of Figure 4.

Looking at Figure 2, we might expect $\mathrm{fdr}_A(z)$, the local fdr for the front, to be much lower than the combined $\mathrm{fdr}(z)$ for large values of $z$, that is, to provide many more "significant" $z$-values, but this is not the case: in fact,

$$(3.5) \qquad \widehat{R}_A(z) = \widehat{\pi}_{A0}(z) / \widehat{\pi}_A(z) \doteq 0.94$$

for $z \geq 3.0$, so formula (3.1) implies only small differences. Two contradictory effects are at work: the longer right tail of the front-half distribution by itself would produce small values of $R_A(z)$ and $\mathrm{fdr}_A(z)$; however, this effect is mostly canceled by the rightward shift of the whole front-half distribution, which substantially increases the numerator of $\mathrm{fdr}_A(z) = p_{A0} f_{A0}(z) / f_A(z)$, (2.11). [Note: the "significant" voxels in Figure 2 were obtained using the theoretical null (1.1) for both the separate and combined analyses, making them somewhat different than those based on the empirical null estimates here.]

The close match between $\widehat{\pi}_{A0}(z)$ and $\widehat{\pi}_A(z)$ near $z = 0$ is no accident. Following through the definitions in Figure 3 and (2.11) gives, after a little rearrangement,

$$(3.6) \qquad \frac{\pi_A(z)}{1 - \pi_A(z)} = \frac{\pi_{A0}(z)}{1 - \pi_{A0}(z)} \frac{1 + Q_A(z)}{1 + Q_B(z)},$$

where

$$(3.7) \qquad Q_A(z) = \frac{1 - \mathrm{fdr}_A(z)}{\mathrm{fdr}_A(z)} \quad \text{and} \quad Q_B(z) = \frac{1 - \mathrm{fdr}_B(z)}{\mathrm{fdr}_B(z)}.$$

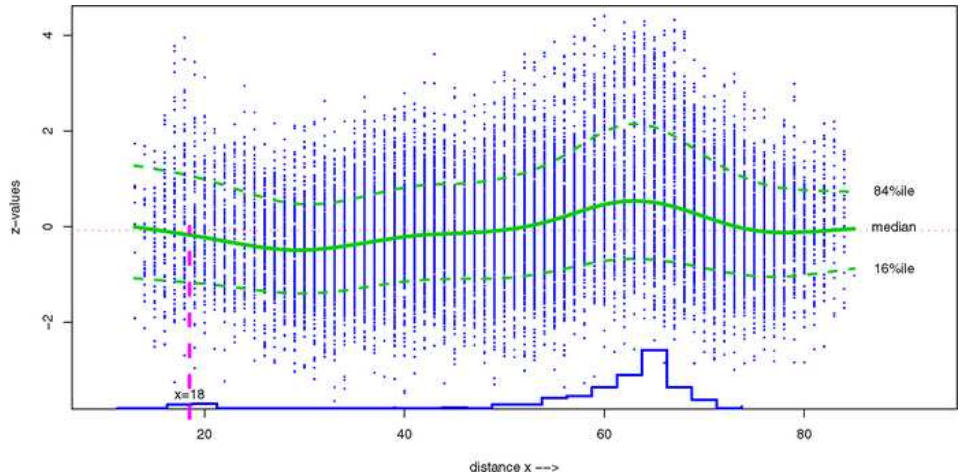

Fig. 5. *Brain Data: z-values plotted vertically versus distance x from back of brain. Small histogram shows x values for the 198 voxels with $z_i \geq 3.02$, left panel of Figure 2. Running percentile curves reveal general upward shift of z-values near histogram mode at $x = 65$.*

Usually $\text{fdr}_A(z)$ and $\text{fdr}_B(z)$ will approximately equal 1.0 near $z = 0$, reflecting the large preponderance of null cases, (2.3), and the fact that nonnull cases tend to produce z-values further away from 0. Then (3.6) gives

$$(3.8) \qquad \pi_A(z) \doteq \pi_{A0}(z) \qquad \text{for } z \text{ near } 0,$$

as seen in Figure 4.

Suppose we believe that $f_{A0}(z) = f_{B0}(z)$ in Figure 3, in other words, that the null cases are distributed identically in the two classes. [This being true, e.g., if we accept the theoretical null distribution (1.1), as is usual in the microarray literature.] Then $\pi_{A0}(z)$ will be constant as a function of $z$,

$$(3.9) \qquad \pi_{A0}(z) = \frac{\pi_A p_{A0}}{\pi_A p_{A0} + \pi_B p_{B0}} = \frac{\pi_A p_{A0}}{p_0}.$$

Since $\hat{\pi}_A(z)$ in Figure 4 is *not* constant near $z = 0$, and should closely approximate $\pi_{A0}(z)$ there, we have evidence against $f_{A0}(z) = f_{B0}(z)$ in this case, obtained without recourse to models such as (3.3).

**4. Fdr estimation for small subclasses.** Our division of the Brain Data into front and back halves was somewhat arbitrary. Figure 5 shows the $N = 15443$ z-values plotted versus $x$, the distance from the back of the brain. A clear wave is visible, cresting near $x = 65$. Most of the 198 $BH(0.1)$ significant voxels of Figure 2 occurred at the top of the crest.

There is something worrisome here: the z-values near $x = 65$ are shifted upward across their entire range, not just in the upper percentiles. This might be due to a reading bias in the DTI imaging device, or a genuine difference between dyslexic and normal children for *all* brain locations around $x = 65$. In neither case is it correct to assign some special significance to those voxels near $x = 65$ having large $z_i$ scores. In fact, a separate Fdr analysis of the 1956 voxels having $x$ in the range $[60, 69]$ [using *locfdr* to estimate their empirical null distribution as $N(0.65, 1.44^2)$] yielded *no* significant cases.

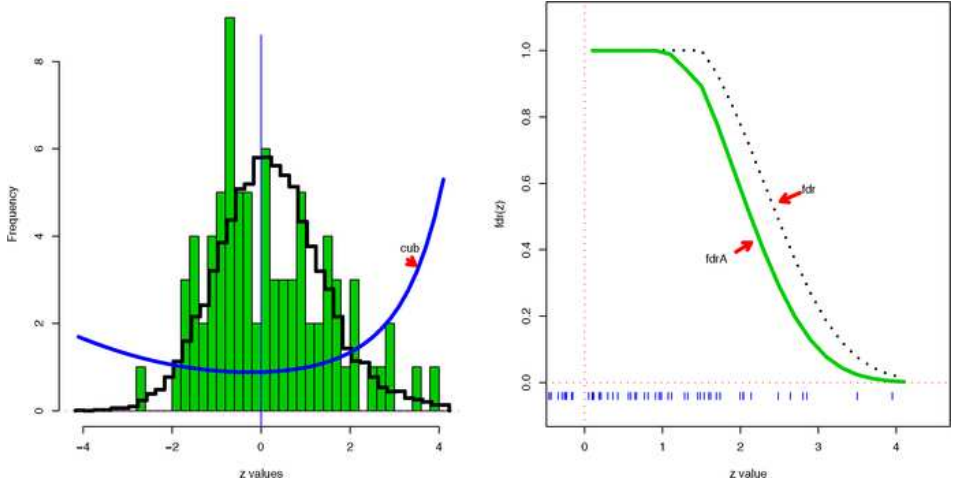

Fig. 6. *Left panel: Solid histogram shows 82 z-values for class A, the voxels at $x = 18$; Line histogram for all other z-values in Brain Data; "cub" is cubic logistic regression estimate of $1/R_A(z)$. Right panel: combined local false discovery rate $\widehat{\text{fdr}}(z)$, obtain from locfdr algorithm, and $\widehat{\text{fdr}}_A(z) = \widehat{\text{fdr}}(z) \cdot \widehat{R}_A(z)$. Dashes indicate the 82 $z_i$'s.*

Figure 5 suggests that we might wish to perform separate analyses on many small subclasses of the data. Large classes can be investigated directly, as above, but a fully separate analysis may be too much to ask for a subclass of less than a few hundred cases, as shown by the accuracy calculations of Efron (2007b). Relationship (3.1) can be useful in such situations.

As an example, let class A be the 82 voxels located at $x = 18$. These display some large $z$-values in Figure 5, attained without the benefit of riding a wave crest. Figure 6 shows their $z$-value histogram and a cubic logistic regression estimate of $\pi_A(z)/\pi_A$, where

$$(4.1) \qquad \pi_A = 82/15543 = 0.0053.$$

This amounts to an estimate of $1/R_A(z)$ in (3.1). Here we are assuming that A shares a common null distribution with all the other cases, $f_{A0}(z) = f_{B0}(z)$ as in (3.9), a necessary assumption since there isn't enough data in A to separately estimate $\pi_{A0}(z)$. In its favor is the flatness of $\widehat{\pi}_A(z)/\pi_A$ near $z = 0$, a necessary diagnostic signal as discussed at the end of Section 3. [Remark

G of Section 8 discusses replacing $\pi_A$ with $\pi_{A0}$, (3.9), in the estimation of $R_A(z)$.]

The right panel of Figure 6 compares the combined local false discovery rate $\widehat{\text{fdr}}(z)$, estimated using *locfdr* as in Efron (2007a), with

$$(4.2) \qquad \widehat{\text{fdr}}_A(z) = \widehat{\text{fdr}}(z) \cdot \widehat{R}_A(z),$$

for $z \geq 0$. The adjustment is substantial. Whether or not it is genuine depends on the accuracy of $\widehat{R}_A(z)$, as considered further in the "efficiency" discussion of Section 7.

So far we have only discussed the dichotomization of cases into two classes A and B of possibly separate relevance. Figure 5 might suggest a more continuous approach in which the relevance of case $j$ to case $i$ falls off smoothly as $|x_j - x_i|$ increases, for example, as $1/(1 + |x_j - x_i|/10)$. We suppose that each case comprises three components,

$$(4.3) \qquad \text{case}_i = (x_i, I_i, z_i),$$

where $x_i$ is an observable vector of covariates, $I_i$ is the unobservable null indicator having $I_i = 1$ or 0 as $\text{case}_i$ is from the null or nonnull group in (2.1), and $z_i$ is the observable $z$-value. We also define a "relevance function" $\rho_i(x)$ taking values in the interval $[0, 1]$, which says how relevant a case with covariate $x$ is to $\text{case}_i$, the case of interest. [Previously, $\rho_i(x_j) = 1$ or 0 as $x_j$ was or was not in the same class as $x_i$.]

The Two-Class model of Figure 3 can be extended to a multi-class model, where each covariate value $x$ has its own distribution parameters $p_{x0}, f_{x0}(z)$, and $f_{x1}(z)$, giving

$$(4.4) \qquad \begin{aligned} f_x(z) &= p_{x0}f_{x0}(z) + (1 - p_{x0})f_{x1}(z) \qquad \text{and} \\ \text{fdr}_x(z) &= p_{x0}f_{x0}(z)/f_x(z) \end{aligned}$$

as in (2.11). Let $\text{fdr}(z)$ be the combined local false discovery rate as in the Theorem of Section 2, and $\text{fdr}_i(z)$ the separate rate $\text{fdr}_{x_i}(z)$. Then (3.1) generalizes to

$$(4.5) \qquad \text{fdr}_i(z) = \text{fdr}(z) \cdot R_i(z), \qquad R_i(z) = \frac{E_0\{\rho_i(x)|z\}}{E\{\rho_i(x)|z\}},$$

"$E_0$" indicating null case conditional expectation.

Tail area false discovery rates (2.5) also satisfy (4.5) after the requisite definitional changes,

$$(4.6) \qquad \text{Fdr}_i(z) = \text{Fdr}(z) \cdot R_i(z), \qquad R_i(z) = \frac{E_0\{\rho_i(X)|Z \leq z\}}{E\{\rho_i(X)|Z \leq z\}}.$$

The empirical version of (4.6) clarifies its meaning. Let $p_{j0}$ and $F_{j0}(z)$ indicate $p_{x0}$ and the c.d.f. of $f_{x0}(z)$ for $x = x_j$, (4.4), and let $N(z) = \#\{z_j \leq$

$z$}. Taking account of all the different situations (4.3), the combined Fdr estimate (2.5) becomes

$$(4.7) \qquad \overline{\mathrm{Fdr}}(z) = \sum_{j=1}^{N} p_{j0} F_{j0}(z) / N(z),$$

this being the ratio of expected null cases to observed total cases for $z_j \leq z_i$. Similarly,

$$(4.8) \qquad \overline{\mathrm{Fdr}}_i(z) = \sum_{j=1}^{N} \rho_i(x_j) p_{j0} F_{j0}(z) \Big/ \sum_{z_j \leq z} \rho_i(x_j),$$

the ratio of expected null cases to total cases taking account of the relevance of $x_j$ to $x_i$. Therefore,

$$(4.9) \qquad \overline{\mathrm{Fdr}}_i(z) = \overline{\mathrm{Fdr}}(z) \cdot \bar{R}_i(z),$$

$$\bar{R}_i(z) = \frac{\left[\sum_{j=1}^{N} \rho_i(x_j) p_{j0} F_{j0}(z) / \sum_{j=1}^{N} p_{j0} F_{j0}(z)\right]}{\left[\sum_{z_j \leq z} \rho_i(x_j) / N(z)\right]};$$

the denominator of $\bar{R}_i(z)$ is an obvious estimate of $E\{\rho_i(X) | Z \leq z\}$ in (4.6), while the numerator is the Bayes estimate of $E_0\{\rho_i(X) | Z \leq z\}$.

## 5. Are separate analyses legitimate?

The principal, and sometimes only, concern of classical multiple inference theory was the control of Type I error in simultaneous hypothesis testing situations. This raises an important question: is it legitimate from an error control viewpoint to split such a situation into separate analysis classes? The answer, discussed briefly here, depends upon the method of inference.

The Bonferroni method applied to $N$ independent hypothesis tests rejects the null for those cases having $p$-value $p_i$ sufficiently small to account for multiplicity,

$$(5.1) \qquad p_i \leq \alpha/N,$$

$\alpha = 0.05$ being the familiar choice. If we now separate the cases into two classes of size $N/2$, rejecting for $p_i \leq \alpha/(N/2)$, we effectively double $\alpha$. Some adjustment of Bonferroni's method is necessary if we are contemplating separate analyses. Changing $\alpha$ to $\alpha/2$ works here, but things get more complicated for situations such as those suggested by Figure 5.

False discovery rate methods are more forgiving—usually they can be applied to separate analyses in unchanged form without undermining their inferential value. Basically, this is because they are *rates*, and, as such, correctly scale with "$N$." We will discuss both Bayesian and frequentist justifications for this statement.

Starting in a Bayesian framework, as in (4.3), let

$$(5.2) \qquad (X, I, Z)$$

represent a random case, where $X$ is an observed covariate vector, $I$ is an unobserved indicator equaling 1 or 0 as the case is null or nonnull, and $Z$ is the observed $z$-value. We assume $X$ has prior distribution $\pi(x)$. $X$ might indicate class A or B in Figure 3, or the distance from the back of the brain in Figure 5.

Let $\mathrm{Fdr}_x(z)$ be the Bayesian tail area false discovery rate (2.5) conditional on observing $X = x$,

$$(5.3) \qquad \mathrm{Fdr}_x(z) = \mathrm{Prob}\{I = 1 | X = x, \ Z \leq z\}.$$

(Remembering that we could just as well change the $Z$ condition to $Z \geq z$ or $|Z| \leq z$.) For each $x$, define a threshold value $z(x)$ by

$$(5.4) \qquad \mathrm{Fdr}_x(z(x)) = q$$

for some preselected control level $q$, perhaps $q = 0.10$. This implies a rule $\mathcal{R}$ that makes decisions "$\hat{I}$" according to (5.4),

$$(5.5) \qquad \hat{I} = \begin{cases} 1 \ (\text{null}), & \text{if } Z > z(X), \\ 0 \ (\text{nonnull}), & \text{if } Z \leq z. \end{cases}$$

Rule $\mathcal{R}$ has a conditional Bayesian false discovery rate $q$ for every choice of $x$,

$$(5.6) \qquad \mathrm{Fdr}_x(\mathcal{R}) \equiv \mathrm{Prob}\{I = 1 | \hat{I} = 0, \ X = x\} = q.$$

Unconditionally, the rate is also $q$:

$$
\begin{aligned}
(5.7) \qquad \mathrm{Fdr}(\mathcal{R}) &= \mathrm{Prob}\{I = 1 | \hat{I} = 0\} \\
&= \int_{\mathcal{X}} \mathrm{Prob}\{I = 1 | Z \leq z(X), X = x\} \pi(x | Z \leq z) \, dx \\
&= \int_{\mathcal{X}} q \cdot \pi(x | Z \leq z) \, dx = q.
\end{aligned}
$$

This verifies the Bayesian separation property for false discovery rates: $\mathrm{Fdr}_x(\mathcal{R}) = q$ separately for all $x$ implies $\mathrm{Fdr}(\mathcal{R}) = q$. Separating the Fdr analyses has not weakened the Fdr interpretation for the entire ensemble.

For any fixed value of $z$, the combined Bayesian false discovery rate $\mathrm{Fdr}(z)$ (2.5) is an a posteriori mixture of the separate $\mathrm{Fdr}_x(z)$ values,

$$(5.8) \qquad \mathrm{Fdr}(z) = \int_{\mathcal{X}} \mathrm{Fdr}_x(z) \pi(x | Z \leq z) \, dx,$$

by the same argument as in the top line of (5.7). There will be some threshold value "$z(\text{comb})$" that makes

$$(5.9) \qquad \mathrm{Fdr}(z(\text{comb})) = q,$$

defining a combined decision rule $\mathcal{R}_{\text{comb}}$ that, like $\mathcal{R}$, controls the Bayes false discovery rate at $q$. Because of (5.8), $z(\text{comb})$ will lie within the range of $z(X)$; $\mathcal{R}_{\text{comb}}$ will be more or less conservative than the separated rule $\mathcal{R}$ as $z(\text{comb}) < z(X)$ or $z(\text{comb}) > z(X)$.[1] Not using the information in $X$ reduces the theoretical accuracy of rule $\mathcal{R}_{\text{comb}}$; see Remark A of Section 8.

Result (5.7) justifies Fdr separation from a Bayesian point of view. The corresponding frequentist/empirical Bayes calculations lead to essentially the same conclusion, though not as cleanly as in (5.7).

In order to justify the empirical Bayes interpretation of $\overline{\text{Fdr}}(z)$ (2.6), we would like it to accurately estimate $\text{Fdr}(z)$ (2.5). Following model (2.1), define

$$(5.10) \qquad N(z) = \#\{z_i \le z\} \quad \text{and} \quad e(z) = E\{N(z)\} = N \cdot F(z);$$

also let

$$(5.11) \qquad D = \frac{\overline{\text{Fdr}}(z)}{\text{Fdr}(z)} \quad \text{and} \quad d = \frac{1 - F(z)}{e(z)} = \frac{1}{N} \frac{1 - F(z)}{F(z)}.$$

Assuming independence of the $z_i$ (just for this calculation) standard binomial results show that

$$(5.12) \qquad E\{D\} \doteq 1 + d \quad \text{and} \quad \text{var}\{D\} \doteq d,$$

where the approximations are accurate to order $O(1/N)$, ignoring terms $O(1/N^2)$. Remark B improves (5.12) to accuracy $O(1/N^2)$.

We see that $\overline{\text{Fdr}}(z)$ is nearly unbiased for $\text{Fdr}(z)$ with coefficient of variation

$$(5.13) \qquad \text{CV}(\overline{Fdr}(z)) = d^{1/2} \le e(z)^{-1/2}.$$

We need $e(z)$ to be reasonably large to make CV small enough for accurate estimation, perhaps

$$(5.14) \qquad e(z) = N \cdot F(z) \ge 10 \qquad \text{for CV} \le 0.3.$$

If we are working near the 1% tail of $F(z)$, common enough in Fdr applications, we need $N \ge 1000$.

Making "$N$" as large as possible in the best reason for combined rather than separate analysis, at least in an empirical Bayes framework. The separate analyses still have large $N$'s in the front-back example of Figure 2, but not in Figure 6. Section 7 shows how the small subclass approach used in Section 4 can improve estimation efficiency.

---

[1] Genovese, Roeder and Wasserman (2006) consider more qualitative situations where what we have called "A" or "B" might be classes of greater or less a priori null probability. Their "weighted BH" rule transforms $z$ into values $z_A$ or $z_B$ depending upon the class, and then carries out rule (2.5) on the transformed $z$'s. Here, instead, the $z$'s are kept the same, but compared to different thresholds $z(x)$. Ferkinstad et al. (2007) explore the dependence of $\text{Fdr}(z)$ on $x$ via explicit parametric models.

The danger of combination is that we may be getting a good estimate of the wrong quantity: if $\text{Fdr}_A(z)$ is much different than $\text{Fdr}_B(z)$, $\overline{\text{Fdr}}(z)$ may be a poor estimate of both.

Returning to a combined analysis, let $N_0(z)$ and $N_1(z)$ be the number of null and nonnull $z_i$'s equal or less than $z$, for some fixed value of $z$,

$$(5.15) \quad N_o(z) = \#\{z_i \leq z, I_i = 1\} \quad \text{and} \quad N_1(z) = \#\{z_i \leq z, I_i = 0\}$$

in notation (4.3), so $N_0(z) + N_1(z) = N(z)$, (5.10); also let $e_0(z)$ and $e_1(z)$ be their expectations,

$$(5.16) \qquad \begin{aligned} e_0(z) &= \text{E}\{N_0(z)\} = Np_0 \cdot F_0(z) \quad \text{and} \\ e_1(z) &= \text{E}\{N_1(z)\} = Np_1 \cdot F_1(z). \end{aligned}$$

The rule that declares $\widehat{I}_i = 0$ if $z_i \leq z$ (i.e., "rejects the null" for $z_i \leq z$) has actual *false discovery proportion*

$$(5.17) \qquad \text{Fdp}(z) = \frac{N_0(z)}{N_0(z) + N_1(z)}.$$

Fdp $(z)$ is unobservable, but we can estimate it by $\overline{\text{Fdr}}(z)$ (2.2), equaling $e_0(z) / (N_0(z) + N_1(z))$ in notation (5.15). This is conservative in the frequentest sense of being an upwardly biased estimate. In fact, it is upwardly biased given any fixed value of $N_1(z)$:

$$(5.18) \qquad \begin{aligned} E\{\overline{\text{Fdr}}(z) \mid N_1(z)\} &= E\left\{\frac{e_0(z)}{N_0(z) + N_1(z)}\Big| N_1(z)\right\} \\ &\geq \frac{e_0(z)}{e_0(z) + N_1(z)} \\ &\geq E\left\{\frac{N_0(z)}{N_0(z) + N_1(z)}\Big| N_1(z)\right\} \\ &= E\{\text{Fdp}(z) \mid N_1(z)\}, \end{aligned}$$

where Jensen's inequality has been used twice. Only the definition $E\{N_0(z) = e_0(z)\}$ is required here, not independent $z_i$'s.

Now suppose we have separated the cases into classes A and B, employing separate rejection rules $z_i \leq z_A$ and $z_i \leq z_B$ satisfying (in the obvious notation)

$$(5.19) \qquad E\{\overline{\text{Fdr}}_A(z_A)\} = E\{\overline{\text{Fdr}}_B(z_B)\} \equiv q.$$

Applying (5.18) shows that the separate false discovery proportions will be controlled in expectation at rate $q$. However, for the equivalent of the

Bayesian result (5.7) to hold frequentistically, we want the combined False Discovery Proportion,

$$(5.20) \qquad \text{Fdp}_{\text{comb}} = \frac{N_A(z_A) + N_{B0}(z_B)}{N_{A0}(z_A) + N_{B0}(z_B) + N_{A1}(z_A) + N_{B1}(z_B)},$$

to satisfy $E\{\text{Fdp}_{\text{comb}}\} \leq q$. Remarks C and D show that asymptotically

$$(5.21) \qquad E\{\text{Fdp}_{\text{comb}}\} = q - \frac{c}{N} + O(1/N^2)$$

for some $c \geq 0$, but an exact finite-sample result has not been verified. (It fails if the denominator expectations are very small.)

Simulations show the original Benjamini–Hochberg rule behaving in the same way; applying rule (2.8) separately to classes A and B also controls the overall expected value of $\text{Fdp}_{\text{comb}}$ at rate $q$, in the sense of (5.20). But again this has not been verified analytically.

The conclusion of this section is that separate false discovery rates analyses are legimate, in the sense that they do not inflate the combined Fdr control rate, at least not if the denominator expectations are reasonably large.

## 6. Enrichment calculations.

Microarray studies frequently yield disappointing results because of low power for detecting individually significant genes, Efron (2007a). "Enrichment" techniques strive for increased power by pooling the $z$-values from some pre-identified collection of genes, for instance, those from a specified pathway, as in Subramanian et al. (2005), Newton et al. (2007) and Efron and Tibshirani (2007). By thinking of the pooled collection as "class A" in (2.16), the Theorem of Section 2 can be brought to bear on enrichment analysis.

Figure 7 involves a microarray study of 10,100 genes, featured in Subramanian et al. (2005), concerning transcription factor, p53. The study compared 17 normal cell lines with 33 lines exhibiting p53 mutations. Two-sample $t$-tests yielded $z$-values $z_i$ for each gene, but the results were disappointing: a standard Fdr test (2.8), with $q = 0.1$, yielded only one nonnull gene, "BAX."

The solid histogram in the left panel of Figure 7 shows $z_i$ values for the 40 genes in set "P53_UP," a collection of genes known to be up-regulated by gene p53. Compared with the line histogram of the 10,060 other $z_i$'s, the P53_UP set definitely looks "enriched," even though it contains only one individually significant $z$-value.

The same analysis as in Figure 6 was applied in Figure 7, with class A now the 40 P53_UP genes. The right panel shows $\widehat{\text{Fdr}}(z)$ for the combined analysis of all 10,100 genes [obtained from *locfdr*, using the theoretical null (1.1)], and

$$(6.1) \qquad \widehat{\text{fdr}}_A(z) = \widehat{\text{fdr}}(z) \cdot \frac{\pi_A}{\hat{\pi}_A(z)},$$

with $\pi_A = 40/10,100$ as in (4.1) and $\hat{\pi}_A(z)$ obtained from a cubic logistic regression. We see that $\widehat{\text{fdr}}_A(z)$ is much smaller than $\widehat{\text{fdr}}(z)$ for $z \geq 2$. Six of the P53_UP genes have $\widehat{\text{fdr}}_A(z_i) < 0.10$, now indicating strong evidence of being nonnull.

The null hypothesis of no enrichment can be started as $\text{fdr}_A(z) = \text{fdr}(z)$. Assuming $\pi_{A0}(z)$ constant, as in (3.9), Theorem (2.16) provides the equivalent statement

(6.2)             *enrichment null hypothesis* : $\pi_A(z) = \text{constant}$,

so we can use $\pi_A(z)$ to test for enrichment. For instance, we might estimate $\pi_A(z)$ with a linear logistic regression, and use the test statistic

(6.3)                                 $S = \hat{\beta}/\widehat{se}$,

where $\hat{\beta}$ is the estimated regression slope and $\widehat{se}$ its estimated standard error.

For the P53_UP gene set, (6.3) gave $S = 4.54$, two-sided $p$-value $6.10^{-6}$. This agrees with the analyses in Subramanian et al. and Efron and Tibshirani, both of which judged P53_UP "enriched," even taking account of simultaneous testing for several hundred other gene sets. (In situations like that of Figure 6, where the class A $z_i$'s extend across a wide range of negative and positive values, $S$ should be calculated separately for $z < 0$ and $z > 0$; in Figure 6 the positive $z_i$'s yielded $S = 3.23, p$-value 0.001.)

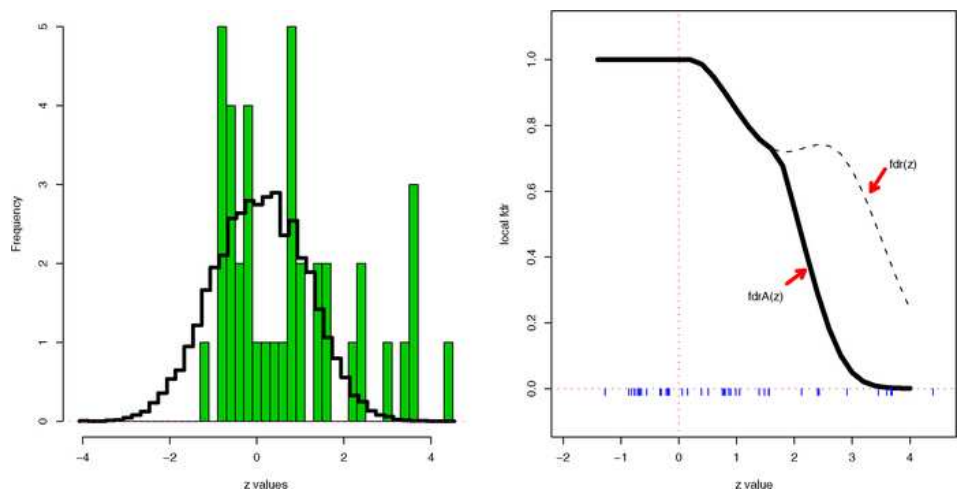

Fig. 7. *p53 microarray study, 10,000 genes, comparing normal versus mutated cell lines. Solid histogram in left panel shows z-values for 40 genes in class P53_UP, compared with all others (line histogram). Right panel compares* $\widehat{\text{fdr}}(z)$ *for all 10,000 genes with* $\widehat{\text{fdr}}_A(z)$ *obtained as in (4.12). Six of the P53_UP genes have* $\widehat{\text{fdr}}_A(z) < 0.1$.

Remark E connects (6.3) to more familiar enrichment test statistics, and suggests that it is likely to be reasonably efficient. The approach here has one notable advantage: we obtain an assessment of individual significance for the genes within the set, via $\widehat{\mathrm{fdr}}_A(z_i)$, rather than just an overall decision of enrichment.

**7. Efficiency.** We used the Theorem of Section 2 to estimate $\mathrm{fdr}_A(z)$ for subclass A:

$$(7.1) \qquad \widehat{\mathrm{fdr}}_A(z) = \widehat{\mathrm{fdr}}(z) \frac{\pi_A}{\hat{\pi}_A(z)},$$

[setting $\pi_{A0}(z)$ in (2.6) equal to $\pi_A$ in Figures 6 and 7]. Of course, we could also estimate, $\mathrm{fdr}_A(z)$ or the tail area analogue $\mathrm{Fdr}_A(z)$, (2.19), directly from the class A data alone, but (7.1) is substantially more efficient. This section gives a very brief overview of the efficiency calculation, with Remark E of Section 8 providing a little more detail, concluding with a simulation example supporting the accuracy of our methodology.

Taking logarithms in (7.1) gives

$$(7.2) \qquad \log \widehat{\mathrm{fdr}}_A(z) = \log \widehat{\mathrm{fdr}}(z) + \log \hat{R}_A(z), \qquad [\hat{R}_A(z) = \pi_A / \hat{\pi}_A(z)].$$

It turns out that $\log \widehat{\mathrm{fdr}}(z)$ and $\log \hat{R}_A(z)$ are nearly uncorrelated with each other, leading to a convenient approximation for the standard deviation of $\log \widehat{\mathrm{fdr}}_A(z)$:

$$(7.3) \qquad sd\{\log \widehat{\mathrm{fdr}}_A(z)\} = [(sd\{\log \widehat{\mathrm{fdr}}(z)\})^2 + (sd\{\log \hat{R}_A(z)\})^2]^{1/2}.$$

Section 5 of Efron (2007b) provides an accurate delta method formula for $sd\{\log \widehat{\mathrm{fdr}}(z)\}$; $sd\{\log \hat{R}_A(z)\}$ is also easy to approximate, using familiar logistic regression calculations.

We expect $\widehat{\mathrm{fdr}}_A(z)$ to be more variable than $\widehat{\mathrm{fdr}}(z)$ since class A involves only proportion $\pi_A$ of all $N$ cases; standard sample-size considerations imply

$$(7.4) \qquad sd\{\log \widehat{\mathrm{fdr}}_A(z)\} \sim \frac{1}{\sqrt{\pi_A}} \, sd\{\log \widehat{\mathrm{fdr}}(z)\}$$

if $\widehat{\mathrm{fdr}}_A(z)$ and $\widehat{\mathrm{fdr}}(z)$ were estimated directly. Estimation method (7.1) does better—the extra variability added to $\widehat{\mathrm{fdr}}(z)$ by $\hat{R}_A(z) = \pi_A / \hat{\pi}_A(z)$, represented by the last term in (7.3), tends to be much smaller than (7.4) suggests.

Figure 8 illustrates a simulation example. In terms of the Two-Class model of Figure 3, the parameters are

$$(7.5) \qquad \begin{aligned} &N = 5000, \qquad \pi_A = 0.01, \qquad p_{A0} = 0.5, \qquad p_{B0} = 1.0 \\ &f_{A0} = f_{B0} \sim N(0, 1) \quad \text{and} \quad f_{A1} = \sim N(2.5, 1); \end{aligned}$$

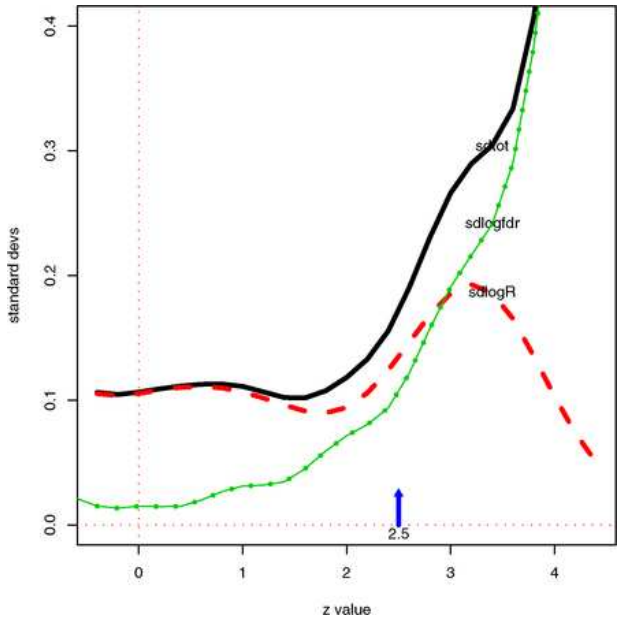

Fig. 8.   *The three standard deviation terms of* (7.3) *for simulation model* (7.5).

so we expect 50 of the 5000 $z_i$'s to be from class A, and 25 of these to be nonnulls distributed as $N(2.5, 1)$, the remaining 4975 cases being $N(0, 1)$ nulls.

At $z = 2.5$, the center of the nonnull distribution, the ratio

$$sd\{\log\widehat{\text{fdr}}_A(z)\}/sd\{\log\widehat{\text{fdr}}(z)\}$$

is 1.61, compared to the ratio 10 suggested by (7.4). Similar results were obtained for other choices of the simulation parameters, (7.5). See Remark H.

**8. Remarks.**   The remarks of this section expand on some of the technical points raised earlier.

REMARK A (*Information loss if X is ignored*).   The observed covariate $X$ in (5.2), or $x_i$ in (4.3), is an ancillary statistic that affects the posterior probability of a null case, $\text{fdr}_x(z) = \text{Prob}\{I = 1 | X = x, Z = z\}$. General principles say that ignoring $X$ will increase prediction error for $I$, and this can be made precise by considering specific loss functions.

Suppose we wish to predict a binary variate $I$ that equals 1 or 0 with probability $p$ or $1 - p$; for a prediction "$P$" in $(0, 1)$, let the loss function be

$$(8.1) \qquad\qquad Q(I, P) = q(P) + \dot{q}(P)(I - P),$$

where $q(\cdot)$ is a positive concave function on $(0, 1)$ satisfying $q(0) = q(1) = 0$ [e.g., $q(p) = p \cdot (1 - p)$ or $q(p) = -\{p \log(p) + (1 - p) \log(1 - p)\}$], and $\dot{q}(p) =$ $dq/dp$. This is the "Q class" of loss functions discussed in Efron (2004a). It turns out that choosing $P$ equal to the true probability $p$ minimizes expected loss, with risk $E\{Q(I, p)\} = q(p)$.

Given $Z = z$, the marginal false discovery rate is

$$(8.2) \qquad \mathrm{fdr}(z) = \mathrm{Prob}\{I = 1 \mid Z = z\} = \int_{\mathcal{X}} \mathrm{fdr}_x(z)\pi(x|z)\, dx,$$

similar to (5.8). Then, with $\mathrm{fdr}(z)$ or $\mathrm{fdr}_x(z)$ playing the role of the true probability $P$, we have

$$(8.3) \qquad q(\mathrm{fdr}(z)) = q\left(\int_{\mathcal{X}} \mathrm{fdr}_x(z)\pi(x|z)\, dx\right) \geq \int_{\mathcal{X}} q(\mathrm{fdr}_x(z))\pi(x|z)\, dx$$

by Jensen's inequality—in other words, the unconditional marginal risk $q(\mathrm{fdr}(z))$ exceeds the expected risk conditioning on $x$.

REMARK B [*Coefficient of variation of* $\overline{\mathrm{Fdr}}(z)$]. Standard calculations involving the first three moments of a binomial variate yield the mean and variance of $D = \overline{\mathrm{Fdr}}(z)/\mathrm{Fdr}(z)$,

$$(8.4) \qquad D \overset{\cdot}{\sim} (1 + d - d^2 c, \; d - d^2(6c - 1))$$

for $d = (1 - F(z))/(N \cdot F(z))$ as in (5.11), and $c = (1 - 2F(z))/(1 - F(z))$, with errors $O(1/N^3)$. This gives approximate coefficient of variation

$$(8.5) \qquad CV(\overline{\mathrm{Fdr}}) \doteq d^{1/2}[1 - d(3c - 1/2)],$$

improving on (5.13).

REMARK C (*Poisson model for* Fdr *relationship*). Let $\mathcal{Z}$ indicate some region of interest in the space of $z$-values for Figure 3, for instance $z \leq z_A$ in the $A$ branch and $z \leq z_B$ in the $B$ branch. Denote the number of null, nonnull, and total $z_i$ values in $\mathcal{Z}$ as $N_0(\mathcal{Z})$, $N_1(\mathcal{Z})$ and $N(\mathcal{Z}) = N_0(\mathcal{Z}) + N_1(\mathcal{Z})$, with corresponding expectations $e_0(\mathcal{Z})$, $e_1(\mathcal{Z})$ and $e(\mathcal{Z}) = e_0(\mathcal{Z}) + e_1(\mathcal{Z})$. Section 5 considers the relationship of three quantities,

$$(8.6) \qquad \overline{\mathrm{Fdr}}(\mathcal{Z}) = \frac{e_0(\mathcal{Z})}{N(\mathcal{Z})}, \qquad \mathrm{Fdr}(\mathcal{Z}) = \frac{e_0(\mathcal{Z})}{e(\mathcal{Z})} \quad \text{and} \quad \mathrm{Fdp}(\mathcal{Z}) = \frac{N_0(\mathcal{Z})}{N(\mathcal{Z})},$$

the estimated Benjamini–Hochberg FDR, the Bayesian Fdr and the False Discovery Proportion.

Now assume that $N$ in Figure 3 is Poisson with expectation $\mu$, and that the $z_i$'s are independent,

$$(8.7) \qquad N \sim \mathrm{Poi}(\mu), \qquad z_1, z_2, \ldots, z_N \text{ independent},$$

implying that $N_0(\mathcal{Z})$ and $N_1(\mathcal{Z})$ are independent Poisson variates,

$$(8.8) \qquad N_0(\mathcal{Z}) \sim \mathrm{Poi}(e_0(\mathcal{Z})) \text{ independent of } N_1(\mathcal{Z}) \sim \mathrm{Poi}(e_1(\mathcal{Z})).$$

We can write

$$(8.9) \qquad N_0(\mathcal{Z}) = e_0(\mathcal{Z}) + \delta_0,$$

where $\delta_0$ has first three moments

$$(8.10) \qquad \delta_0 \sim (0, e_0(\mathcal{Z}), e_0(\mathcal{Z})),$$

and similarly for $N_1(\mathcal{Z}) \equiv e_1(\mathcal{Z}) + \delta_1$ and $N(\mathcal{Z}) \equiv e(\mathcal{Z}) + \delta$.

NOTE. The independence in (8.7) is not a necessary assumption, but it leads to the neatly specific forms of the relationship below.

The Poisson assumptions make it easy to relate the two random quantities $\overline{\mathrm{Fdr}}(\mathcal{Z})$ and $\mathrm{Fdp}(\mathcal{Z})$ in (8.6) to the parameter $\mathrm{Fdr}(\mathcal{Z})$:

LEMMA. *Under assumption (8.7),*

$$(8.11) \qquad E\{\overline{\mathrm{Fdr}}(\mathcal{Z})\} \doteq \mathrm{Fdr}(\mathcal{Z}) \cdot (1 + 1/e(\mathcal{Z})) + O(1/e(\mathcal{Z})^2)$$

*and*

$$(8.12) \qquad E\{\mathrm{Fdp}(\mathcal{Z})\} \doteq \mathrm{Fdr}(\mathcal{Z}) + O(1/e(\mathcal{Z})^2).$$

*Typically, $e(\mathcal{Z}) = e_0(\mathcal{Z}) + e_1(\mathcal{Z})$ will be $O(\mu)$, so that the error terms in (8.11)–(8.12) are $O(1/\mu^2)$, effectively $O(1/N^2)$. The Lemma shows that $\mathrm{Fdr}(\mathcal{Z})$, the Bayesian false discovery rate, is an excellent approximation to $E\{\mathrm{Fdp}(\mathcal{Z})\}$, while $E\{\overline{\mathrm{Fdr}}(\mathcal{Z})\}$ is only slightly upwardly biased.*

PROOF. Following through definitions (8.6) and (8.9),

$$
\begin{aligned}
(8.13) \qquad \overline{\mathrm{Fdr}}(\mathcal{Z}) - \mathrm{Fdr}(\mathcal{Z}) &= \frac{e_0(\mathcal{Z})}{e(\mathcal{Z}) + \delta} - \frac{e_0(\mathcal{Z})}{e(\mathcal{Z})} \\
&= \mathrm{Fdr}(\mathcal{Z}) \left[ \frac{1}{1 + \delta/e(\mathcal{Z})} - 1 \right] \\
&\doteq \mathrm{Fdr}(\mathcal{Z}) \left[ -\frac{\delta}{e(\mathcal{Z})} + \frac{\delta^2}{e(\mathcal{Z})^2} \right],
\end{aligned}
$$

so taking expectations yields (8.11). Similarly,

$$
\begin{aligned}
(8.14) \qquad \mathrm{Fdr}(\mathcal{Z}) - \mathrm{Fdp}(\mathcal{Z}) &= \mathrm{Fdr}(\mathcal{Z}) \cdot \left[ 1 - \frac{1 + \delta_0/e_0(\mathcal{Z})}{1 + \delta/e(\mathcal{Z})} \right] \\
&\doteq \mathrm{Fdr}(\mathcal{Z}) \cdot \left[ 1 - \left( 1 + \frac{\delta_0}{e_0(\mathcal{Z})} \right) \left( 1 - \delta/e(\mathcal{Z}) + \frac{\delta^2}{e(\mathcal{Z})^2} \right) \right].
\end{aligned}
$$

Therefore,

$$(8.15) \quad \begin{aligned} E\{\mathrm{Fdr}(\mathcal{Z}) - \mathrm{Fdp}(\mathcal{Z})\} &\doteq \mathrm{Fdr}(\mathcal{Z})\left[E\left\{\frac{\delta_0\delta}{e_0(\mathcal{Z})e(\mathcal{Z})}\right\} - E\left\{\frac{\delta^2}{e(\mathcal{Z})^2}\right\}\right] \\ &= \mathrm{Fdr}(\mathcal{Z})\left[\frac{e_0(\mathcal{Z})}{e_0(\mathcal{Z})e(\mathcal{Z})} - \frac{e(\mathcal{Z})}{e(\mathcal{Z})^2}\right] = 0. \qquad \square \end{aligned}$$

REMARK D (*Frequentist* Fdr *combination result*).   The Lemma above leads to a heuristic verification of (5.21): that under (5.19), $E\{\mathrm{Fdr}_{\mathrm{comb}}\} \leq q$ in (5.20). To begin with, notice that (8.7) gives expected values of $N_{A0}(z_A)$ and $N_A(z_A)$,

$$(8.16) \quad e_{A0}(z_A) = \mu\pi_A p_{A0}F_{A0}(z_A) \quad \text{and} \quad e_A(z_A) = \mu\pi_A F_A(z_A),$$

so

$$(8.17) \quad \frac{e_{A0}(z_A)}{e_A(z_A)} = \frac{p_{A0}F_{A0}(z_A)}{F_A(z_A)} = \mathrm{Fdr}_A(z_A),$$

the Bayesian Fdr for class A, and similarly, $e_{B0}(z_B)/e_B(z_B) = \mathrm{Fdr}_B(z_B)$. From (8.11) and (5.19), applied individually within the two classes,

$$(8.18) \quad \begin{aligned} \mathrm{Fdr}_A(z_A) &\doteq q\left[1 - \frac{1}{e_A(z_A)}\right] \leq q \quad \text{and} \\ \mathrm{Fdr}_B(z_B) &\doteq q\left[1 - \frac{1}{e_B(z_B)}\right] \leq q. \end{aligned}$$

Therefore, the combined Bayesian Fdr is also bounded by $q$,

$$(8.19) \quad \begin{aligned} \mathrm{Fdr}_{\mathrm{comb}} &= \frac{e_{A0}(z_A) + e_{B0}(z_B)}{e_A(z_A) + e_B(z_B)} \\ &= \frac{e_A(z_A)\,\mathrm{Fdr}_A(z_A) + e_B(z_B)\,\mathrm{Fdr}_B(z_B)}{e_A(z_A) + e_B(z_B)} \\ &\leq \frac{e_A(z_A)q + e_B(z_B)q}{e_A(z_A) + e_B(z_B)} = q. \end{aligned}$$

But $E\{\mathrm{Fdp}_{\mathrm{comb}}\} \doteq \mathrm{Fdr}_{\mathrm{comb}}$ according to (8.12), verifying (5.21).

REMARK E (*The slope statistic for testing enrichment*).   Slope statistic (6.3), $S = \hat{\beta}/\widehat{se}$, is asymptotically fully efficient for enrichment testing under a two-sample exponential families model. Suppose that all the $z$-values come, independently, from a one-parameter exponential family having density functions

$$(8.20) \quad g_\eta(z) = e^{\eta z - \psi(\eta)}g_0(z),$$

as in Lehmann and Romano (2005), with

$$(8.21) \qquad \eta = \begin{cases} \eta_A, & \text{in class A,} \\ \eta_B, & \text{in class B.} \end{cases}$$

Define $\beta = \eta_A - \eta_B$. As $N \to \infty$, the MLE $\hat{\beta}$ has asymptotic null hypothesis distribution

$$(8.22) \qquad \hat{\beta} \overset{\cdot}{\sim} N\left(\beta, \ \frac{1}{N \pi_A \pi_B V(\eta_0)}\right),$$

where $\pi_A$ and $\pi_B$ are the proportions of $z_i$'s from the two classes, and $V(\eta_0)$ is the variance of $z$ if $\eta_A = \eta_B$ equals, say, $\eta_0$.

Bayes rule applied to (8.20)–(8.21) gives

$$(8.23) \quad \text{logit}(\pi_A(z)) = \beta z + c \qquad \left[c = \log\left(\frac{\pi_A}{\pi_B}\right) + \psi(\eta_B) - \psi(\eta_A)\right],$$

with $\beta = \eta_A - \eta_B$ as above. Standard calculations show that $\hat{\beta}$ obtained from the logistic regression model (8.23) also satisfies (8.22). This implies full asymptotic efficiency of the slope statistic (6.3) for testing $\eta_A = \eta_B$, the "no enrichment" null hypothesis. Under normal assumptions, $g_\eta(x) \sim N(\eta, 1)$ in (8.20), (6.3) is asymptotically equivalent to $\bar{z}_A - \bar{z}_B$, the difference of class means; "limma" [Smyth (2004)], an enrichment test implemented in *Bioconductor*, is also based on $\bar{z}_A$, as discussed in Efron and Tibshirani (2007b).

REMARK F [*The additive variance approximation* (7.3)]. Being a little more careful, we can use (3.9) to write (7.1) as

$$(8.24) \qquad \widehat{\text{fdr}}_A(z) = \widehat{\text{fdr}}(z)\frac{\pi_{A0}}{\hat{\pi}_A(z)} \qquad \left(\pi_{A0} = \frac{\pi_A p_{A0}}{\pi_A p_{A0} + \pi_B p_{B0}}\right),$$

under the assumption that $f_{A0}(z) = f_{B0}(z)$ in Figure 3. Binning the data as in (3.2) gives

$$(8.25) \qquad \ell\widehat{\text{fdr}}_{Ak} = \ell\widehat{\text{fdr}}_k - \ell\hat{\pi}_{Ak} + \log(\pi_{A0}),$$

where $\ell\widehat{\text{fdr}}_{Ak}$ is $\log(\widehat{\text{fdr}}_A(z))$ evaluated at the midpoint $z_{(k)}$ of bin $k$, and similarly, $\ell\widehat{\text{fdr}}_k = \log(\widehat{\text{fdr}}(z_{(k)}))$ and $\ell\hat{\pi}_{Ak} = \log(\hat{\pi}_A(z_{(k)}))$. *Locfdr* computes the estimates $\ell\widehat{\text{fdr}}_k$ from the vector of counts $\mathbf{N} = (\dots, N_k, \dots)$, while a standard logistic regression program computes $\ell\hat{\pi}_{Ak}$ from the vector of proportions $\mathbf{r}_A = (\dots, r_{Ak} = N_{Ak}/N_k, \dots)$, (3.2); $\mathbf{N}$ and $\mathbf{r}_A$ are, to a first order of calculation, uncorrelated, leading to approximation (7.3).

In broad outline, the argument depends on the general equality

$$(8.26) \qquad \text{var}\{X + Y\} = \text{var}\{X\} + \text{var}\{Y\} + 2\,\text{cov}\{X, E(Y|X)\},$$

applied to $X = \ell \widehat{\text{fdr}}_k$ and $Y = -\ell \hat{\pi}_{Ak}$. Both $\text{var}\{X\}$ and $\text{var}\{Y\}$ are $O(1/N)$, but, because the expectation of $\mathbf{r}_A$ does not depend on $\mathbf{N}$, the covariance term in (8.26) is of order only $O(1/N^2)$, and can be ignored in (7.3).

REMARK G (*The assumption of identical null distributions*).   If we are willing to assume that $f_{A0}(z) = f_{B0}(z)$ in Figure 3 [or equivalently, that $f_{A0}(z) = f_0(z)$, the combined null density], then relationship (3.1) becomes

$$(8.27) \qquad \text{fdr}_A(z) = \text{fdr}(z)\frac{\pi_{A0}}{\pi_A(z)} \qquad [\pi_{A0} = \pi_A p_{A0}/p_0],$$

as in (3.9), which can be written as

$$(8.28) \qquad \text{fdr}_A(z) = \text{fdr}(z)\frac{\pi_A}{\pi_A(z)}\frac{p_{A0}}{p_0}.$$

The examples in Figures 6 and 7 estimated $R_a(z) = \pi_{A0}/\pi_A(z)$ by $\pi_A/\hat{\pi}_A(z)$, ignoring the final factor $p_{A0}/p_0$ in (8.28). This is probably conservative: one would expect a small subclass A of interest to have proportionately more nonnull cases than the whole ensemble, in other words, to have $p_{A0}/p_0 < 1$.

It isn't difficult to estimate the full relationship (8.27). Since $\pi_{A0} \doteq \pi_A(z)$ for $z$ near 0, (3.8)–(3.9), we can set

$$(8.29) \qquad \widehat{\text{fdr}}_A(z) = \widehat{\text{fdr}}(z)\frac{\hat{\pi}_A(0)}{\hat{\pi}_A(z)};$$

(8.29) gives better results if the logistic regression model for estimating $\pi_A(z)$ incorporates a flat interval around $z(0)$—for instance, if only positive values of $z$ are of interest,

$$(8.30) \qquad \text{logit}(\pi_A(z)) = \beta_1 + \beta_2 \max(z-1,0)^2 + \beta_3 \max(z-1,0)^3.$$

REMARK H (*Simulation example*).   Figure 9 graphs 100 simulations of $\widehat{\text{fdr}}_A(z)$, (7.1), drawn from model (7.5). The comparison with the actual curve $\text{fdr}_A(z)$ shows excellent accuracy. Here neither the simulations nor the actual curve incorporate the factor $p_{A0}/p_0$ in (8.28), which could be included as in (8.29)–(8.30).

A simpler correction starts with $\widehat{\text{fdr}}_A(z)$, (7.1), estimates $p_{A0}$ by

$$(8.31) \qquad \hat{p}_{A0} = \sum_A \widehat{\text{fdr}}_A(z_i)/N_A,$$

and finally multiplies $\widehat{\text{fdr}}_A(z_i)$ by $\hat{p}_{A0}/p_0$. In the simulations for Figure 9, $\hat{p}_{A0}$ had mean 0.575 and standard deviation 0.035, reasonably close to the true value $p_{A0} = 0.50$.

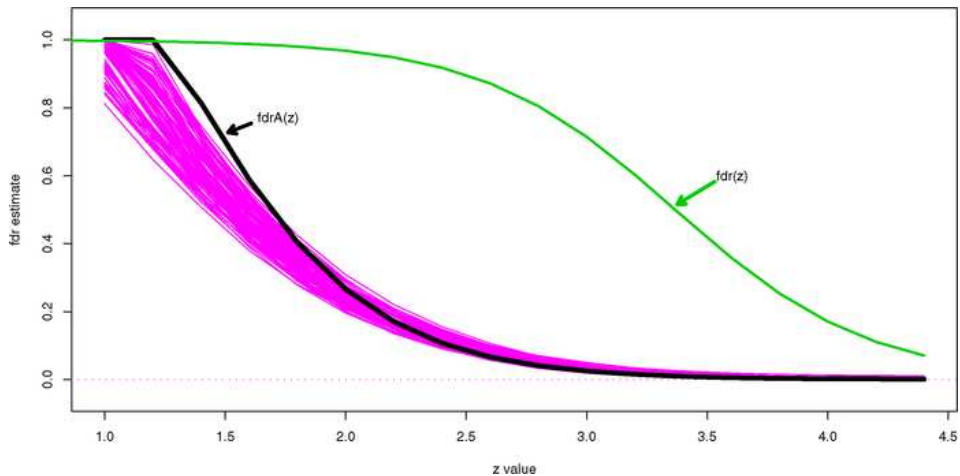

FIG. 9.  *Light lines show* 100 *simulations of* $\widehat{\mathrm{fdr}}_A(z)$, (7.1), *from model* (7.5); *heavy line is actual* $\mathrm{fdr}_A(z)$ *curve. [Factor* $p_{A0}/p_0$ *in* (8.28) *not included in actual or simulations.] Also shown is combined rate* $\mathrm{fdr}(z)$. *Formula* (7.1) *provides good accuracy in this case.*

**9. Discussion and summary.**  A more accurate title for this paper might have been "When *shouldn't* hypothesis testing problems be combined?" A general algorithm for combining or separating problems is beyond my scope here, but the analysis makes it clear that combination can be dangerous in situations like that of Figure 5. On the positive side, the simple Bayesian theorem of Section 2, extended at (4.5), helps signal if separation is called for, and even how it can be efficiently carried out. Some specific points:

- Combining problems increases the empirical Bayes inferential accuracy, with $N > 1000$ necessary for reasonably accurate direct estimation of false discovery rates (5.13)–(5.14), at least in the partially nonparametric framework of Benjamini and Hochberg (1995).
- However, the Separate-Class model of Figure 2, and its ensuing theorem (2.16), imply that separate inferences can be necessary for problems of differing structure. The question of whether to combine problems amounts, here, to a question of trading off variance with bias in the estimation of false discovery rates.
- Situations like that of Figure 5 argue strongly against a single combined analysis. The theorem can be implemented as in Figure 6 to estimate fdr or Fdr for small subclasses, $N = 82$ in Figure 6, and with surprising accuracy as shown in Section 7.
- A formal test for separation can be based on the slope statistic (6.3). This provided strong evidence for the necessity of separation in the p53 enrichment example of Figure 7, and moderately strong evidence in Figure 6.

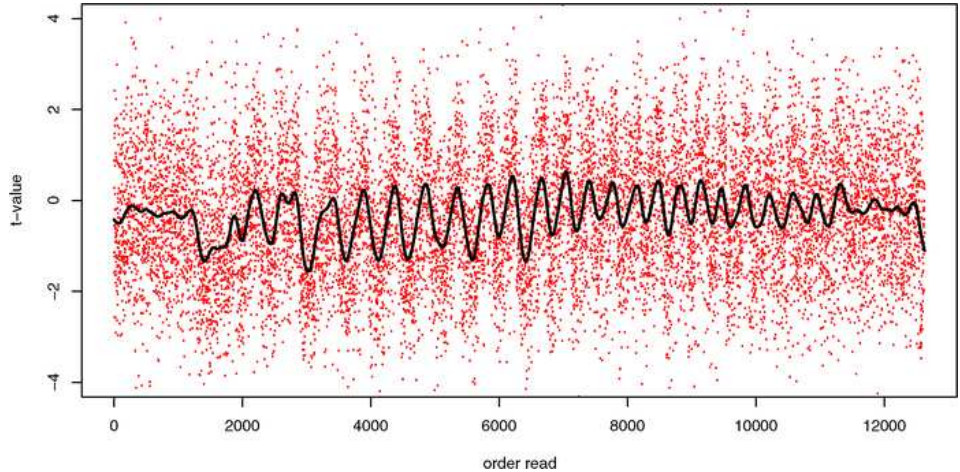

Fig. 10. *Paired t-statistics comparing affected versus nonaffected tissue in 13 cancer patients; microarray study of 12625 genes. The t-values are plotted vertically, against the order in which they were read from the array. Smoothing spline (solid curve) reveals periodic disturbances.*

- Section 5 shows that controlling the false discovery rate in separate classes also controls it in combination, at least if the expected number of tail events isn't too small. In this sense, Fdr analysis has an advantage over other simultaneous testing techniques.

Whether or not the specific methodology presented here appeals to the reader, the general question of which problems to combine in a simultaneous testing situation remains important. As a matter of due diligence, plotting test statistics versus possible covariates, as in Figure 5, can raise a warning flag against casual combination. Such covariates exist even in loosely structured microarray studies—where, for example, the order in which the expression levels are read off the plate can reveal noticeable effects.

This last point is illustrated in Figure 10, where a periodic disturbance in the microarray reading mechanism has evidently affected the gene-wise summary statistics. Subtracting the estimated disturbance function from the observed $t$-statistics is an obviously wise first step. Adjustments that make cases more comparable are a complementary tactic to separate analyses. Both can be useful in large-scale testing situations. In Figure 5, for example, we might adjust the $z$-values by subtracting the local median and dividing by the local spread (84%–16%). The resulting version of Figure 5, however, still displays obvious inhomogeneity, and requires separate analyses like that in Figure 7 to ferret out the interesting cases.

DEPARTMENTS OF STATISTICS
    AND HEALTH, RESEARCH AND POLICY
SEQUOIA HALL
STANFORD UNIVERSITY
STANFORD, CALIFORNIA 94305-4065
USA
E-MAIL: brad@stat.stanford.edu